# Efficient Electrocatalytic Reduction of $CO_2$ by Nitrogen-Doped Nanoporous Carbon/Carbon Nanotube Membranes – A Step Towards the Electrochemical $CO_2$ Refinery

Hong Wang[a], Jia Jia[a], Pengfei Song[b], Qiang Wang[c], Debao Li[c], Shixiong Min[d], Chenxi Qian[a], Lu Wang[a], Young Feng Li[a], Chun Ma[e], Tom Wu[e], Jiayin Yuan*[f,g], Markus Antonietti[f], Geoffrey A. Ozin*[a]

**Abstract:** The search for earth abundant, efficient and stable electrocatalysts that can enable the chemical reduction of $CO_2$ to value-added chemicals and fuels at an industrially relevant scale, is a high priority for the development of a global network of renewable energy conversion and storage systems that can meaningfully impact greenhouse gas induced climate change. Here we introduce a straightforward, low cost, scalable and technologically relevant method to manufacture an all-carbon, electroactive, nitrogen-doped nanoporous carbon-carbon nanotube composite membrane, dubbed "HNCM/CNT". The membrane is demonstrated to function as a binder-free, high-performance electrode for the electrocatalytic reduction of $CO_2$ to formate. The Faradaic efficiency for the production of formate is 81%. Furthermore, the robust structural and electrochemical properties of the membrane endow it with excellent long-term stability.

Sustainable conversion of greenhouse gas carbon dioxide ($CO_2$) into value-added products is the subject of extensive research because of the ever-increasing global $CO_2$-levels and environmental concerns.[1-4] The electrocatalytic $CO_2$ reduction reaction ($CO_2$RR) conducted in aqueous media is extremely appealing in terms of multiple merits, which include the use of the abundant and benign solvent water, the operation under ambient temperature and pressure conditions, and the implementation by applying an electric potential and therefore a means of storing excess renewable electricity off the grid.[5,6] Achieving this goal requires efficacious electrocatalysts, since $CO_2$ is fully oxidized and stable.

Searching for suitable electrocatalysts for the $CO_2$RR, represents one of the most active areas of current research. Known catalysts for this reaction are noble metals (e.g. Au, Ir, Ag, Pd),[7-13] base metals, alloys and their oxides (e.g. Cu, Sn, $Cu_2O$, $SnO_2$, $CoO_x$),[14-21] 2D metal dichalcogenides (e.g. $MoS_2$,[22] $WSe_2$[23]), covalent organic frameworks (COF),[24] metal organic frameworks (MOFs),[25,26] homogeneous molecular catalysts,[27,28] and N-doped carbons.[29-36]

With respect to the practical implementation of $CO_2$RR in aqueous media, existing electrocatalysts suffer from one or more of the following problems: poor selectivity and competitive hydrogen evolution side reaction; low electrochemical stability; difficulty of reclaiming powder form electrocatalysts, complicated synthesis and fabrication processes, high-cost and poisoning of noble metals and the engineering challenges of industrial scale electrochemical reduction of gases.

Among various $CO_2$RR electrocatalysts, N-doped porous carbon materials are particularly attractive due to their low cost, large surface areas, metal-free nature, chemical stability as well as tunable conductivity and electrochemical activity.[37-46] More importantly, the competing $H_2$ evolution reaction by N-doped carbons in water is sluggish and kinetically negligible under appropriate conditions.[47] Regrettably, N-doped carbon electrocatalysts developed so far are almost entirely powder-based and suffer from a low productivity for the $CO_2$RR. These powders have to be engineered into pre-defined shapes by mixing and pressing with electronically insulating polymer binders, such as Nafion or PVDF, to make them useful as practical electrodes. This process, though technologically mature, deteriorates the overall cell electrical conductivity and the contact between catalyst and electrolyte. In addition, owing to mechanically weak contacts between the catalyst, electrode and binder, the electroactive species can become detached thereby reducing the performance and long-term operation of the $CO_2$RR electrode.

To overcome these experimental hurdles and raise the technological potential of N-doped carbon to drive efficient electrocatalytic $CO_2$RR, it would be advantageous to develop hierarchically structured porous N-doped carbon membranes (HNCMs) as binder-free electrodes. In such membranes, the macropores provide mass transport highways while the mesopores and micropores provide a large surface area and high population of spatially accessible electroactive sites for optimal performance in the $CO_2$RR. Furthermore, the nitrogen species incorporated into the carbon framework improve the electrochemical stability and most importantly constitute the active sites for $CO_2$RR. While there have been several notable attempts to produce porous carbon membranes it still remains a challenge to make one of practical significance, due to the difficulty of simultaneously controlling their chemical composition, porosity and structural integrity.

Herein, a nanoporous membrane based on a composite of CNT and HNCM, termed HNCM/CNT, was prepared, which works as a highly active, selective and stable electrode for the aqueous phase $CO_2$RR. The high purity liquid product formate was produced with a Faradaic efficiency as high as 81%. With further research and development this electrochemical $CO_2$RR system could form the platform for a formic acid technology

[a] Dr. H. Wang, Mis. J. Jia, Dr. C. Qian, Dr. L. Wang, Mr. Y. Li, Prof. G. A. Ozin
Materials Chemistry and Nanochemistry Research Group, Solar Fuels Cluster, Centre for Inorganic and Polymeric Nanomaterials, Departments of Chemistry, Chemical Engineering and Applied Chemistry, and Electrical and Computing Engineering, University of Toronto, 80 St. George Street, Toronto, Ontario M5S3H6, Canada
E-mail: gozin@chem.utoronto.ca
[b] Prof. P. Song, College of Chemistry and Chemical Engineering, Northwest Normal University, Lanzhou 730070, China.
[c] Dr. Q. Wang, Dr. D. Li
State Key Laboratory of Coal Conversion, Institute of Coal Chemistry, the Chinese Academy of Sciences Taiyuan 030001, China
[d] Prof. S. Min
School of Chemistry and Chemical Engineering, Beifang University of Nationalities, Yinchuan, Ningxia, China
[e] Mr. C. Ma, Prof. T. Wu
Physical Science and Engineering Division, King Abdullah University of Science & Technology (KAUST), Thuwal, 23955-6900, Saudi Arabia
[f] Prof. J. Yuan, Prof. M. Antonietti
Department of colloid chemistry, Max Planck Institute of Colloids and Interfaces, 14476 Potsdam, Germany
[g] Department of Chemistry and Biomolecular Science and Center for Advanced Materials Processing, Clarkson University, 13699, USA
E mail: jyuan@clarkson.edu

based on a metal-free carbon catalyst.

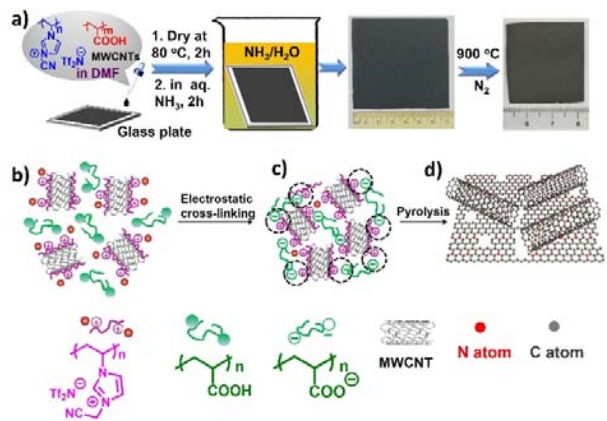

**Scheme 1. a)** scheme illustrating the synthetic route to the membranes. The last two images are photographs of the primary PCMVImTf$_2$N/PAA/CNT film and the final carbon membrane. Scheme of the molecular structure of **b)** homogeneous dispersion of CNTs in a solution of PCMVImTf$_2$N and PAA in DMF; **c)**, a PCMVImTf$_2$N/PAA/CNT film, and **d)** a HNCM/CNT membrane.

The construction was performed *via* a bottom-up method (details was provided in supporting information). **Scheme 1a** shows the synthetic procedure and structural model of the intermediates and the HNCM/CNT hybrid membrane. Firstly, a homogeneous multi-wall CNTs dispersion was prepared by sonicating CNTs in a solution of poly[1-cyanomethyl-3-vinylimidazolium bis(trifluoromethanesulfonyl)imide] (PCMVImTf$_2$N) and poly(acrylic acid) (PAA) in dimethyl formamide (DMF) (**Figure S1**). Owing to the well-known cation-π interactions between the imidazolium cations in PCMVImTf$_2$N and the graphitic CNTs surface,[48] polymer chains attach to the surface of CNTs for dispersion purposes (**Scheme 1b**). Then, the stable polymer/CNTs dispersion was cast onto a glass plate, dried at 80 °C and finally immersed in an aqueous NH$_3$ solution to build up the nanoporous polymer/CNTs film membrane (**Scheme 1c and Figure S2**). Afterwards, pyrolysis of the porous polymer/CNTs film membrane at 900 °C in N$_2$ leads to the targeted HNCM/CNT membrane (**Scheme 1d**). Film thickness can be controlled by the added amount of the polymer/CNTs dispersion. It is worth mentioning that the carbon membranes produced in our laboratory ovens are already several square centimeters in size (**Scheme 1a**). The fabrication method is straightforward and readily adapted to much larger size membranes and their properties rationally optimized using a range of polymeric ionic liquids and carbon nanostructures.

Scanning electron microscope (SEM) image (**Figure 1a**) shows that an asymmetric, three-dimensionally interconnected macropore architecture was created. The pore size can be seen to gradually decrease from 2 to 1.2 μm to 700 nm from the top to the bottom (from zone I and II to III), respectively. A high-magnification SEM image (**Figure 1b and Figure S3**) clearly reveals that CNTs are uniformly embedded in HNCM/CNT (yellow arrows indicate CNTs). High-resolution transmission electron microscopy (HRTEM) provides insight into the microstructure (**Figure 1c**), in which CNTs embedded in a N-doped carbon membrane matrix are observable (white arrows directed area). The well-defined lattice spacing of 0.34 nm indicates that the HNCM/CNT membrane contains highly organized graphitic domains. The single graphitic layers are seen to bend due to the nitrogen doping, but extend practically across the entire membrane. **Figure 1d** shows a HRTEM image of a native CNT, which is typically composed of 7~12 layers with an outer diameter of approximately 5~10 nm. Notably, in **Figure 1e**, a thin rough sheath is formed on the CNT wall (see the area pointed by the yellow arrow), referred to as CNT-NC core-shell microstructure. It has been recently reported that the CNT surface templates the pyrolysis of ionic liquid species on its surface.[49] The newly formed carbon on the surface of CNT is rich in defects and short-range ordered micropores. Specifically, the heteroatoms, through templating interactions, nicely align at the CNTs surface, thereby enabling a high electron transfer efficiency via charge transfer interactions, which in turn will modify catalytic activity. Energy-filtered transmission electron microscopy mappings for both C and N (**Figure 1g-h**) indicate a uniform distribution of N, which is typical for N-rich molecular carbon precursors, in this case, PCMVImTf$_2$N.

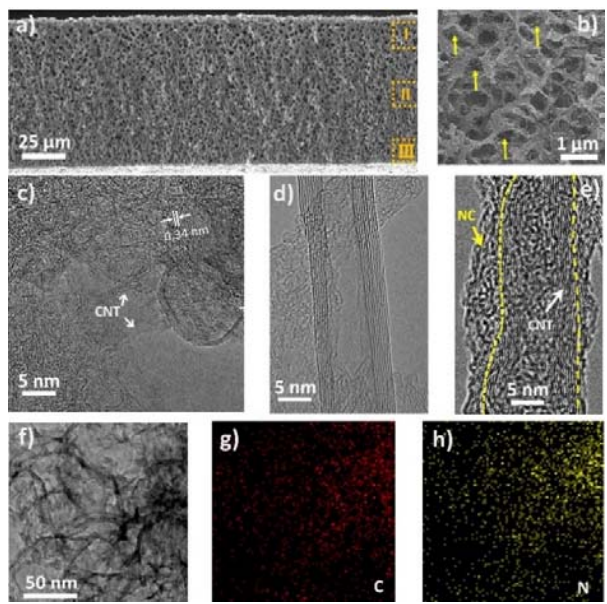

**Figure 1.** Low-magnification **a)** and high-magnification **b)** cross-section SEM images of the membrane; HRTEM images of (**c**) HNCM/CNT hybrid membrane, **d)** native CNTs and **e)** core-shell structure of a CNT-N doped carbon heterojunction. The yellow arrow directed area represents N-doped carbon attached to a CNT marked by a white arrow; **f-h)**, TEM image of the HNCM/CNT hybrid membrane and corresponding elemental mappings.

To understand the intrinsic electrocatalytic properties of the hybrid carbon membrane, we prepared CNT-free N-doped carbon porous membranes (HNCM) as a control reference, and its detailed preparation procedure and structural characterization are provided in Supporting information and **Figure S4**. The N contents and structures of HNCM and HNCM/CNT were analyzed by elemental analysis and X-ray photoelectron spectroscopy (XPS). The elemental analysis show that N content of the HNCM and HNCM/CNT are quite similar, 8.67 at.% and 8.26 at.%, respectively. The XPS spectra (**Figure S5**) show typical C, N and O peaks in HNCM and HNCM/CNT, both being truly metal-free. A thorough analysis of different N species in the HNCM and HNCM/CNT structures are presented in **Figure 2**. The N 1s XPS spectrum of HNCM shows that the N atoms are inserted into the graphene framework mainly in the pyridinic (N1, 398.0 eV), pyrrolic (N2, 398.6 eV) and graphitic (N3, 400.2 eV) forms with abundances of 9.2%, 5.9% and 84.9%, respectively (**Figure 2a**). By contrast, in HNCM/CNT the N atom is mainly existent in the pyridinic and graphitic forms in abundance of 41.8 % and 58.2%, respectively (**Figure 2b**). It is relevant that the pyridinic N content in HNCM/CNT is more than 4 times of that in HNCM, which we attribute to surface templating of the condensation reactions in conjunction with coupled edge termination of graphitic layers by pyridinic units. As a further probe, the intensity ratio of the D/G band (ID/IG) determined in Raman spectroscopy reflects the degree of irregularity in the N-doped carbons.[50] As shown in **Figure S6**, the D band at 1355 cm$^{-1}$ and G band at 1587 cm$^{-1}$ are usually attributed to the defective/disordered structure and ordered carbon structure with long range sp$^2$ electronic configurations for HNCM and HNCM/CNT, respectively. The $I_D/I_G$ values increased from 0.74 for HNCM to 0.91 for HNCM/CNT, reflecting more highly disordered but catalytically more active structures in HNCM/CNT, which is directly related to the increased pyridinic nitrogen content and the coupled edge terminations in HNCM/CNT. Therefore, the incorporation of CNTs appears to favor the preferential formation of pyridinic nitrogens in the porous carbon framework, presumably through topology and electronic interactions. Previous report proposed that the active site of N-doped carbon materials are carbon atoms with Lewis basicity next to pyridinic N[41]. This means that HNCM/CNT has potential high electrocatalytic activity.

Specific surface area plays an essential role in optimizing catalytic activity of heterogeneous catalysts. The Brunauer-Emmett-Teller (BET) specific surface area of HNCM/CNT is calculated to be 546 m$^2$/g, whereas that of HNCM is calculated to be 418 m$^2$/g when prepared under the same conditions (**Figure S7**). The sharp increase of nitrogen sorption at low pressures (P/P$_0$ < 0.05) is due to the nitrogen filling in micropores below 2 nm, quantified by the density functional theory (DFT) pore size distribution curves derived from the N$_2$ adsorption branch (**Figure S8**). The obvious hysteresis above P/P$_0$ ~ 0.5 is indicative of the existence of mesopores. It is clear from the imaging and adsorption studies that the pore architecture of the HNCM/CNT membrane is hierarchical in nature, being comprised of a gradient of macropores seen in the cross-sectional image with pores traversing the micro- to meso- to macropore range. It is noteworthy, in spite of its highly porous nature, the HNCM/CNT membrane displays a high electrical conductivity. As shown in **Figure S9,** the conductivity of the HNCM/CNT membrane is an impressive 134 S cm$^{-2}$ at 298 K, significantly higher than that of HNCM (98 S cm$^{-2}$ at 298 K). The high conductivity of HNCM/CNT favors fast charge transport, a mandatory requirement for efficient electrocatalysis.

In this context, the hierarchical porous architecture of this electrically conductive membrane allows it to function as a diffusion electrode to enhance the three phase contact and charge transport between the electrocatalyst, aqueous electrolyte and gaseous CO$_2$ reactant. This is especially important for optimising the rate of the electrocatalytic reduction of CO$_2$ with its notoriously poor solubility and low concentration in the aqueous phase.

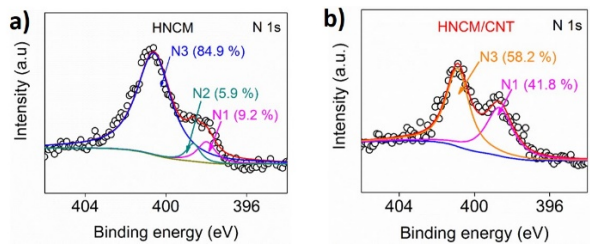

**Figure 2** a). High resolution N1 XPS spectra of b) HNCM and c) HNCM/CNT.

For electrochemical experiments, the HNCM/CNT and HNCM membranes were directly utilized as electrodes without the need for a binder. Cathodic linear sweep voltammetry was first conducted in a CO$_2$-saturated or Argon (Ar)-saturated aqueous 0.1 M KHCO$_3$ solution. As shown in **Figure 3a**, HNCM/CNT exhibits a more positive overpotential and larger reduction current than HNCM electrode, *i.e.* the onset overpotentials of HNCM/CNT and HNCM are -0.18 and -0.37 V, respectively under CO$_2$-saturated aqueous 0.1 M KHCO$_3$ solution, and -0.53 V under Ar-saturated aqueous 0.1 M KHCO$_3$ solution. Clearly, the overpotentials of HNCM/CNT and HNCM electrodes negatively shifted under CO$_2$-saturated aqueous 0.1 M KHCO$_3$ solution compared to that in Ar-saturated aqueous 0.1 M KHCO$_3$ solution, indicating high activity toward CO$_2$RR of our electrodes. It should be noted that HNCM/CNT exhibits more negative overpotential than HNCM in CO$_2$-saturated aqueous 0.1 M KHCO$_3$ solution. This may be due to the preferred binding of HNCM/CNT to CO$_2$ and accelerated kinetics of CO$_2$RR. Comprehensive product analysis using gas chromatography, high performance liquid chromatography and nuclear magnetic resonance (NMR) was carried out to reveal the nature of the chemical processes occurring within our HNCM/CNT and HNCM electrocatalysts. The integrated system was tested for an extended period of 180 minutes (**Figure S10-11**). As illustrated in **Figure 3b** and **Figure S12-13**, the main product is formate, the remainder being a small quantity of CO and H$_2$, which escapes from the aqueous environment into the gas phase for both of HNCM and HNCM/CNT electrodes. The liquid product composition was further confirmed by $^1$H-NMR (**Figure S14**) to be pure formate, supporting the high selectivity of its production.

The electrokinetics of CO$_2$RR at HNCM and HNCM/CNT electrodes were examined using Tafel analysis (**Figure 3c**). Tafel slopes of HNCM and HNCM/CNT are 156 mV/dec and 138 mV/dec in the linear curves, respectively. These values are close to the 118 mV/dec, being consistent with a rate-limiting single-electron transfer process in powdered N-doped carbon.[29]

In addition, the stability of an electrocatalyst is key for practical applications. Both HNCM and HNCM/CNT membranes were operated for 36 h, as shown in **Figure 3d**, and no decay of activity was observed, indicating excellent electrochemical durability for both systems. We conducted XPS analysis of the HNCM/CNT sample after 36 h of electrolysis. As shown in **Figure S15**, similar to the original HNDCM/CNT, the N species in HNCM/CNT after 36 h of electrolysis exist in the form of graphitic N (56%) and pyridinic N (38%), except a slight fraction of oxidized N (6%) produced, indicating its relatively robust electrochemical stability. This feature is very possibly due to the nitrogen doping effect, which improves the electrochemical stability and resistance against oxidation by modifying the electronic band structure of the graphitic carbons.[37]

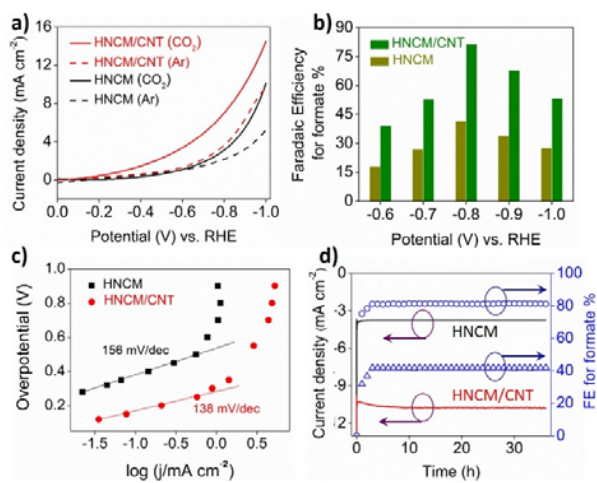

**Figure 3** a), Cathodic linear sweep voltammetry scans at 5 mV/s in a $CO_2$-saturated or argon-saturated aqueous 0.1 M $KHCO_3$ solution. b, Faradaic efficiencies for formate production *vs* applied potential at NCM and NCM/CNTs electrodes. c, Tafel plots of HNCM and HNCM/CNT, respectively. d, Stability of HNCM and HNCM/CNT evaluated through chronoamperometric measurements.

We further compared the least overpotential required for achievable maximum FE of formate for HNDCM/CNT with those reported electrocatalysts in the literatures (**Figure 4**). The achievable maximum FE (81%) of formate for HNDCM/CNT is comparable to that of state-of-the-art metal-free PEI-NCNTs (85%),[29] and higher than that of powderous N-doped graphene.[51] It is noted that the overpotential required to reach the highest FEs of formate is 213 mV lower for HNDCM/CNT (0.9 V) compared to the PEI-NCNT catalyst (1.13 V). In a control experiment, under an optimized condition for HNDCM/CNT (-0.9 V vs RHE), it was found that purified metal-free powderous CNT (SEM image and XPS spectrum were provided in **Figure S16** and **Figure S17**) did not catalyse conversion of $CO_2$ to formate. The formate FE of N-doped carbon derived from CNT/PCMVImTf$_2$N (**Figure S18-19**) was 51%, close to previous reported value for N-doped CNT.[29] However, it should be noted that the steady current density of the powderous N-doped carbon gradually decreased as a function of time (**Figure S20**), indicating that the powderous N-doped carbons alone are not stable during $CO_2$RR under the same condition for HNDCM/CNT. Moreover, we also compared the least overpotentials required for prevalent metal electrodes and nanostructured metal catalysts for achievable maximum FE of formate in the literature studies. It is noteworthy that HNDCM/CNT as metal-free electrocatalysts is in fact one of the best electrocatalysts reported so far for electrolysis reduction of $CO_2$-to-formate.

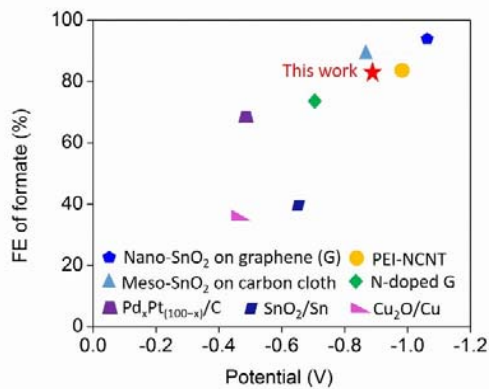

**Figure 4.** Comparison of potentials and FEs of formation of formate on HNDCM/CNT with other electrocatalysts reported in the recently literatures: nano-$SnO_2$[16]; PEI-NCNT[29]; N-doped graphene[51]; meso-$SnO_2$/carbon cloth[52]; $Pd_xPt_{(100-x)}$/C[53]; Sn/$SnO_2$[54]; $Cu_2O$/Cu[55].

The increase of the amount of pyridinic N in N-doped carbon is conducive to improving $CO_2$RR activity and selectivity for the formation of formate in aqueous solution. Based on this experimental results and previous report,[56] we propose the $CO_2$ to formate reaction mechanism illustrated in **Figure 5**. The $CO_2$ molecule is first adsorbed to the basic carbon atom adjacent to the pyridinic N. Subsequently, the adsorbed $CO_2$ molecule is reduced to the COO* radical. The stabilized COO* is then protonated, the proton source most likely being $HCO_3^-$ since its $pK_a$ value (10.33) is smaller than that of $H_2O$ (15.7). This step is followed by a second and rapid electron-transfer reduction reaction to release formate as the product. Further experiments are needed to unveil the detailed mechanism of the electrocatalysed $CO_2$RR by HNCM/CNT hybrid membranes.

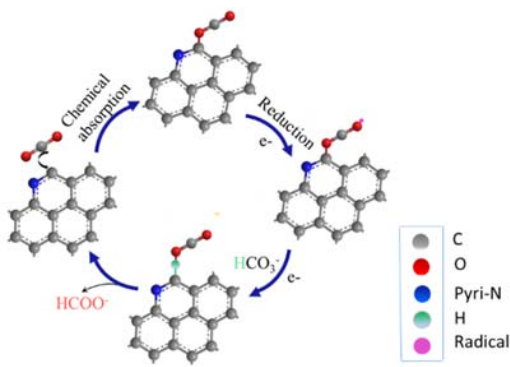

**Figure 5.** Proposed mechanism of the CO$_2$RR in 1M KHCO$_3$ aqueous solution by HNCM/CNT membrane.

In summary, an innovative, simple and easily scalable synthetic strategy to construct nanoporous HNCM/CNT membranes has been developed. Because of their large surface area, excellent electrical conductivity, high population of active centers, hierarchical pore architecture as well as membrane mechanical integrity, these nanoporous HNCM/CNT electrocatalytic membranes function as diffusion electrodes, exhibiting impressive activity and selectivity as well as long term durability for the CO$_2$RR in water. Another unique feature of this genre of poly(ionic liquid)/CNTs membrane is the ability to incorporate any type of metal ion and nanoparticle into the structure.[57] Thereafter, metal nanoparticle functionalized HNCM/CNT membranes can be readily prepared by carbonization of a metal/polyelectrolyte/CNTs membrane. We envision different metal functionalized N-doped carbons will provide myriad opportunities for converting CO$_2$ into desired chemicals and fuels.

## Acknowledgements

G.A.O. is a Government of Canada Research Chair in Materials Chem istry and Nanochemistry. Financial support for this work was provided by the Ontario Ministry of Research Innovation (MRI); Ministry of Economic Development, Employment and Infrastructure (MEDI); Ministry of the Environment and Climate Change; Connaught Innovation Fund; Connaught Global Challenge Fund; and the Natural Sciences and Engineering Research Council of Canada (NSERC). S. M. acknowledges the financial support from the National Natural Science Foundation of China (21463001). H.W. and T.W. also thank the King Abdullah University of Science and Technology (KAUST) for financial support. J. Y. is grateful for financial support from the Max Planck society, Germany, Clarkson University, USA and the ERC (European Research Council) Starting Grant (project number 639720-NAPOLI).

**Keywords:** Carbon nanotubes • Membrane • Nitrogen-doping • Hierarchical porous carbon • CO$_2$ reduction

# Efficient Electrocatalytic Reduction of CO$_2$ by Nitrogen-Doped Nanoporous Carbon/Carbon Nanotube Membranes – A Step Towards the Electrochemical CO$_2$ Refinery

## Materials and Method

1-Vinylimidazole (Aldrich 99%), 2,2'-azobis(2-methylpropionitrile) (AIBN, 98%), bromoacetonitrile (Aldrich 97%), and bis(trifluoromethane sulfonyl)imide lithium salt (Aldrich 99%) were used as received without further purifications. Dimethyl sulfoxide (DMSO), dimethyl formamide (DMF), methanol, and tetrahydrofuran (THF) were of analytic grade. Potassium bicarbonate (KHCO$_3$) was of analytic grade and purchased from Sigma Aldrich. Poly(acrylic acid) (PAA) MW: 130,000 g/mol, was obtained from Sigma Aldrich. Multi-walled carbon nanotube (≥ 98 %, O.D x I.D x L.D 10 nm ± 1 nm x 4.5 nm ± 0.5 nm x 3 ~6 µm) was purchase from Sigma Aldrich and purified by the following procedure: 300 mg of carbon nanotubes were refluxed in 50 mL of 0.3 M HNO$_3$ for 12 h in a 250 mL flask to remove the catalyst. After cooling down, the CNTs were separated by centrifugation, washed with excess of water several times and freeze-dried. Poly[1-cyanomethyl-3-vinylimidazolium bis(trifluoromethane sulfonyl)imide] (abbreviation: PCMVImTf$_2$N) was synthesized according to reference (*S1*). Nitrogen doped carbon membrane was prepared according to previous report (*S2*). A piece of glass substrate was ultrasonically cleaned in acetone and deionized water, respectively, for 30 min before use.

*Preparation of HNCM/CNT*. A homogeneous CNT dispersion was firstly accomplished by sonicating 0.1 g of purified CNTs in 1g of poly[1-cyanomethyl-3-vinylimidazolium bis(trifluoromethane sulfonyl)imide] (PCMVImTf$_2$N) and 0.18g of poly(acrylic acid) (PAA) solution in 10 mL of dimethyl formamide (DMF) for 4 h. Owing to the strong polarization and cation-π interactions between the imidazolium units in PCMVImTf$_2$N and CNTs, PCMVImTf$_2$N specifically attaches to the surface of the CNTs. Then, centrifugation of the dispersion removed aggregates (undispersed CNTs). The formed homogeneous dark dispersion was cast onto a clean glass plate with an area of 3 x 3 cm$^2$ and dried at 80 °C for 3 h. Then the film was immersed in a 0.1 wt % aqueous NH$_3$ solution for 3 h, following a modified procedure reported previously by our group (*S3*). This process triggers charging of PAA and electrostatic crosslinking of the charged PAA with PCMVImTf$_2$N simultaneously, constructing a stable porous film, which can be readily peeled off from the substrate. Afterwards, pyrolysis of the film at 900 ºC in pure $N_2$ under 1.5 torr. leads to the hierarchically structured N-doped HNCM/CNT hybrid.

*Preparation of N-doped carbon derived from CNT/ PCMVImTf$_2$N.* A homogeneous CNT dispersion was accomplished by sonicating 0.1 g of purified CNTs in 1g of poly[1-cyanomethyl-3-vinylimidazolium bis(trifluoromethane sulfonyl)imide] (PCMVImTf$_2$N) in 10 mL of dimethyl formamide (DMF) for 4 h. Then, centrifugation of the dispersion removed aggregates (undispersed CNTs). The powderous CNT/PCMVImTf$_2$N was obtained by drying the solution. Afterwards, N-doped carbon was obtained by carbonization of CNT/PCMVImTf$_2$N at 900 ºC in pure $N_2$ under 1.5 torr.

*Characterization.* $^1$H-NMR spectra were recorded on a Bruker AVANCE III spectrometer operating at 400 MHz resonance frequency. X-ray photoelectron spectroscopy (XPS) data were collected by an Axis Ultra instrument (Kratos Analytical) under ultrahigh vacuum (<$10^{-8}$ Torr) and by using a monochromatic Al Kα X-ray source. The carbon 1s peak was calibrated at 285 eV and used as an internal standard to compensate for any charging effects. Nitrogen sorption isotherms were measured at -196 °C using a Micromeritics ASAP 2020M and 3020M system. The samples were degassed for 6 h at 200 °C before the measurements. Pore size distribution was calculated by density functional theory (DFT) method. Elemental analyses were obtained from the service of Mikroanalytisches Labor Pascher (Remagen, Germany). A field emission scanning electron microscope (FESEM, FEI Quanta 600FEG) was used to acquire SEM images. Transmission electron microscope (TEM) and high resolution TEM (HRTEM) images were taken on a JEOL JEM-2100F transmission electron microscopy operated at 200 kV.

*Electrochemical measurements* the electrochemical measurements were performed on an electrochemical impedance spectroscopy (EIS) capable channel in a Biologic VMP3 potentiostat. The as-prepared HNCMs or HNCM/CNT hybrids were used as the working electrode. A Pt and an Ag/AgCl (in saturated KCl solution) electrode were used as the counter and reference electrodes, respectively. The electrolyte is 0.1 M KHCO$_3$ solution with pH 6.8. All the applied potentials were converted to reversible hydrogen electrode (RHE) potential scale (without the IR compensation) using E (vs. RHE)=E(vs. Ag/AgCl)+ 0.210 V+0.0592 V*6.8. A fixed volume of 30 mL electrolyte was used for all of the electrochemical experiments. For controlled potential electrolysis of $CO_2$, the cathodic compartment of the cell was degassed and saturated with $CO_2$ at 10 mL min$^{-1}$ for 30 minutes. During the electrolysis, $CO_2$ gas was continuously bubbled into the cathodic compartment at a rate of 10 mL min$^{-1}$ and was delivered directly to the sampling loop of an on-line pre-calibrated gas chromatograph (Agilent7890B) ($H_2$, CO, $CH_4$, $C_2H_4$, $C_2H_6$, $CO_2$) (*S4*). A GC run was initiated every 20 min. The gaseous products were analyzed using a packed 13 x molecular sieves with a thermal conductivity detector using He as a carrier gas for $CO_2$, CO, $CH_4$, $C_2H_4$, and $C_2H_6$ analysis. Argon was used as the carrier gas for analysis. The Faradaic efficiencies (FEs) of CO and $H_2$ production were calculated from the volume concentration of gaseous products as below:

$$FE_j(\%) = \frac{2Fv_j Gtp_0}{RT_0 Q_{total} \times 1000000} \times 100\%$$

Where, $v_j$ (vol%) = volume concentration of j = CO or $H_2$ in the exhaust gas from the electrochemical cell at a given sampling time;

G (ml/min at room temperature and ambient pressure) = Gas flow rate;

$Q_{total}$ (C) = Integrated charge passed during electrolysis (Chronoamperometry data);

t (min)= Electrolysis time;

$P_0$ =1.01x10$^5$ Pa, $T_0$ = 269.2 K, F = 96500 C mol$^{-1}$, R=8.314 J mol$^{-1}$ K$^{-1}$;

Liquid product was quantified after the electrochemical measurement by high-pressure liquid chromatography (HPLC, Agilent technologies) system equipped with Agilent 1200, 1260 and 1290 Infinity liquid chromatography technology and further confirmed by $^1$H-NMR (Bruker AVANCE III 400) using $D_2O$ as solvent. The FEs for CO and $H_2$ are average values and that for formate is a cumulative value during the electrolysis. The electrode area was calculated from its surface area.

*Preparation of N-doped carbon derived from CNT/ PCMVImTf$_2$N based Electrode.* A suspension of N-doped carbon powders derived from CNT/PCMVImTf$_2$N and Nafion perfluorinated resin solution along with a mixture of water and isopropanol were extensively mixed by ultrasonicating for 1 hour to produce the catalyst ink, which was then brushed onto the carbon paper, with an approximate 1 × 1 cm$^2$ area. Eventually, a uniform 5 ± 0.2 mg catalyst loading was achieved. The electrodes were then dried at 110 °C overnight in a vacuum oven.

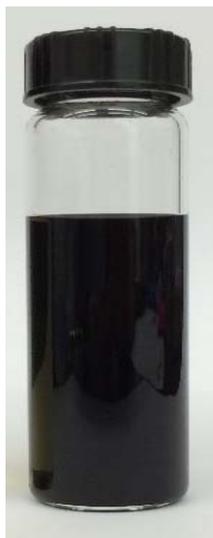

**Figure S1.** Photograph of the homogeneous and stable multi-wall carbon nanotube (CNT) dispersion in poly[1-cyanomethyl-3-vinylimidazolium bis(trifluoromethanesulfonyl)imide] (PCMVImTf$_2$N) and poly(acrylic acid) (PAA) dimethyl formamide (DMF) solution.

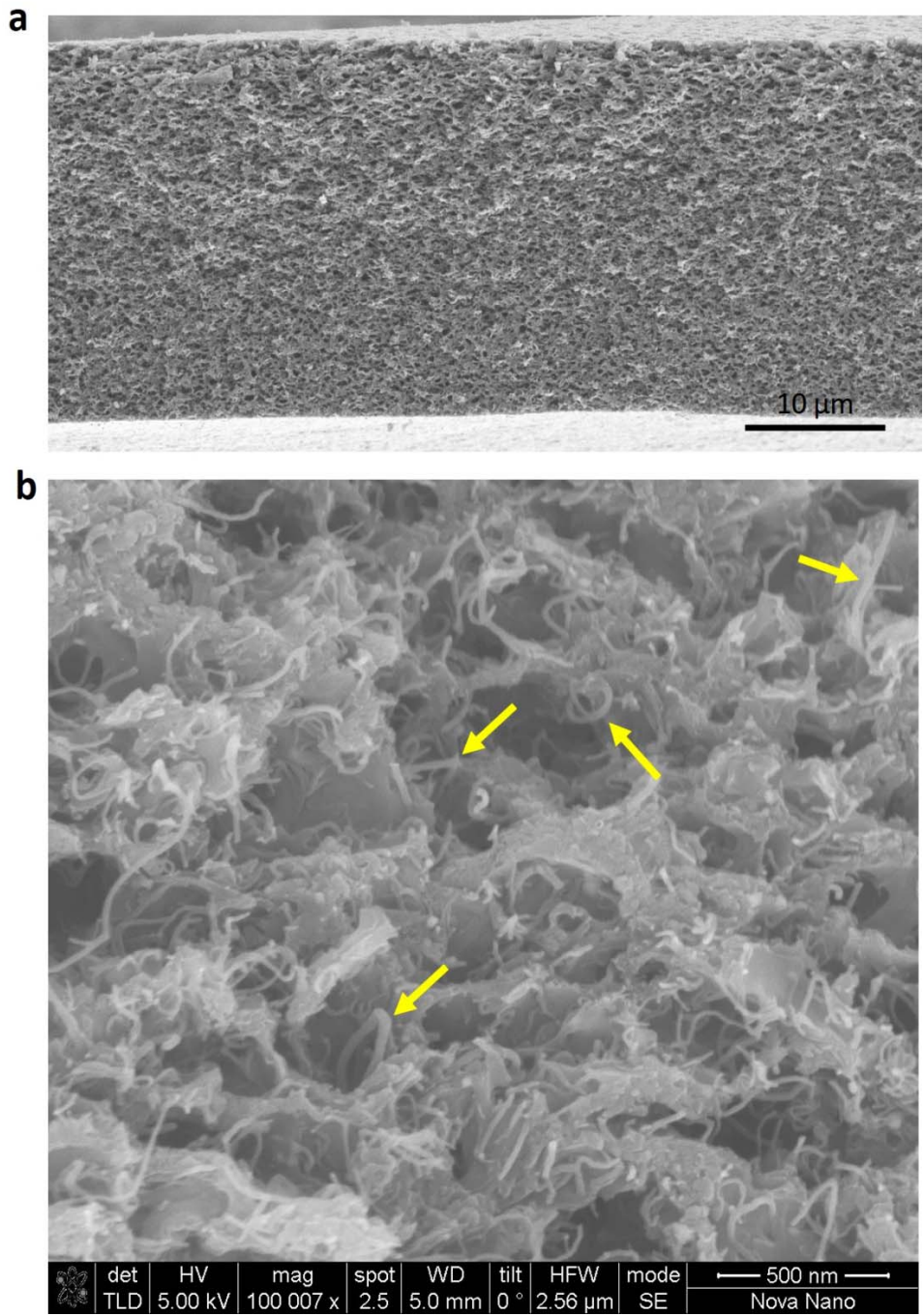

**Figure S2.** Cross-section SEM images of PCMVImTf$_2$N/PAA/CNT membrane. a, Low magnification; b, high magnification. Particularly, from Supplementary Figure 2b, it can be clearly seen that the carbon nanotubes are well dispersed, much thicker than the pristine one (*Multi-walled carbon nanotube (≥ 98 %, O.D x I.D x L.D 10 nm ± 1 nm x 4.5 nm ± 0.5 nm x 3 ~6 μm) was purchase from Sigma Aldrich)*, which can be attributed to the adsorption of PCMVImTf$_2$N on the surface of carbon nanotubes owing to the π-π interactions between the imidazolium cations in PCMVImTf$_2$N and the graphitic CNTs surface.

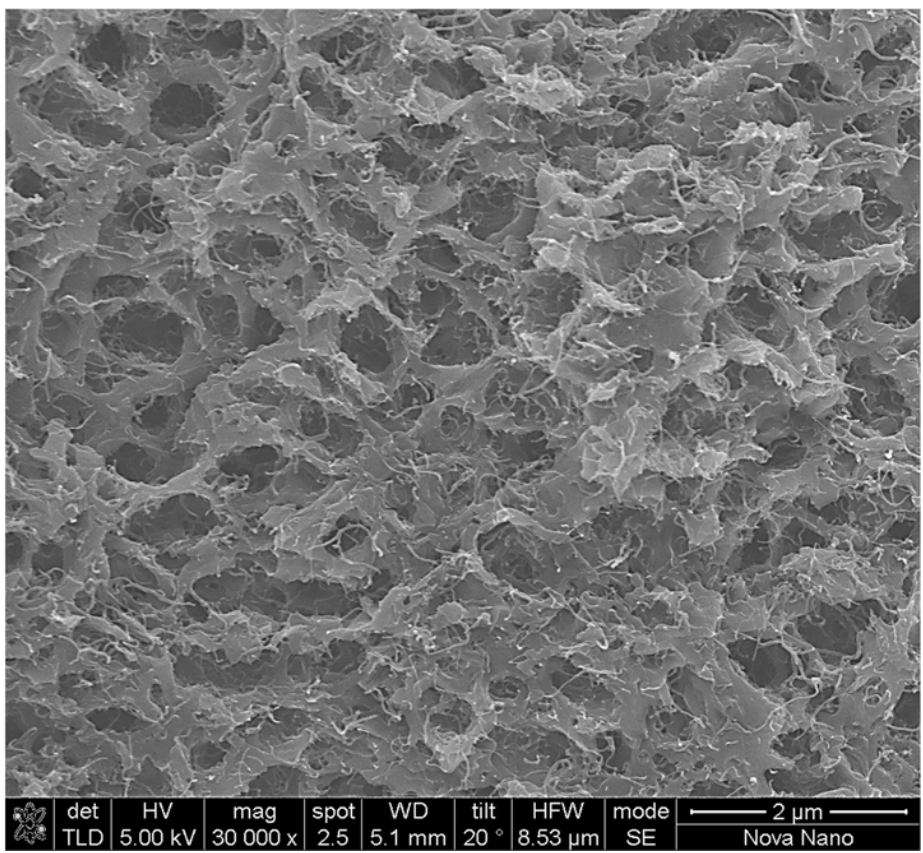

**Figure S3**. Enlarged cross-section SEM image of HNDCM/CNT.

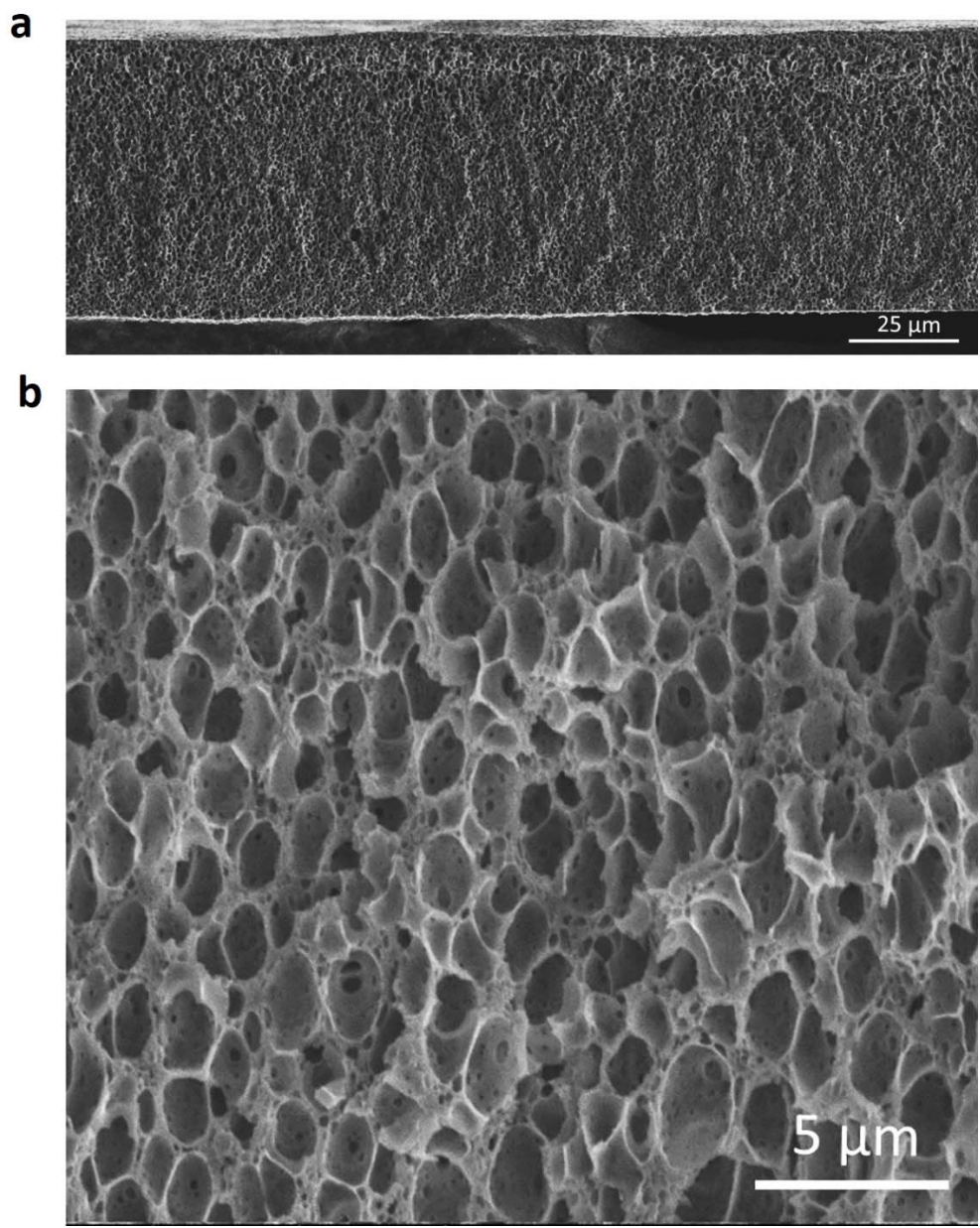

**Figure S4.** Cross-section SEM images of HNCM. a, Low magnification; b, high magnification.

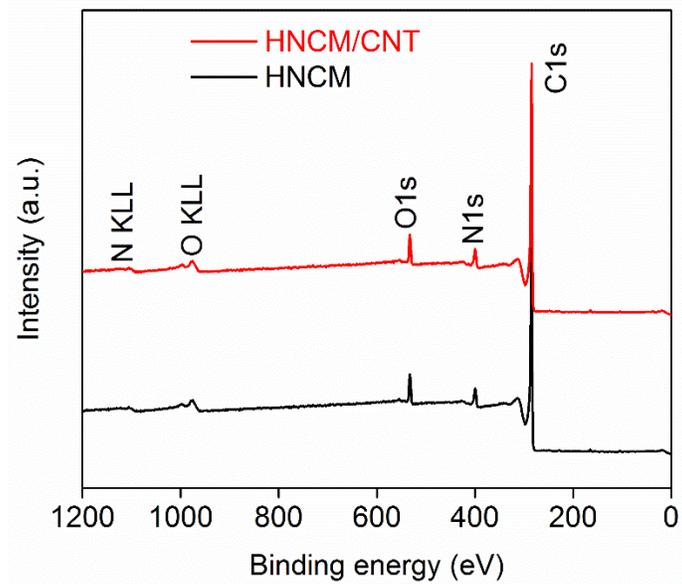

**Figure S5.** XPS spectra of HNCM and HNCM/CNT

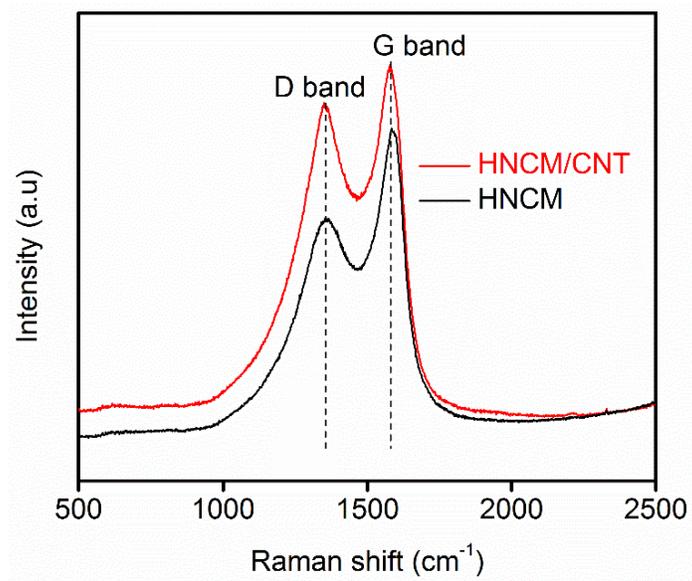

**Figure S6.** Raman spectrum of HNCM and HNCM/CNT.

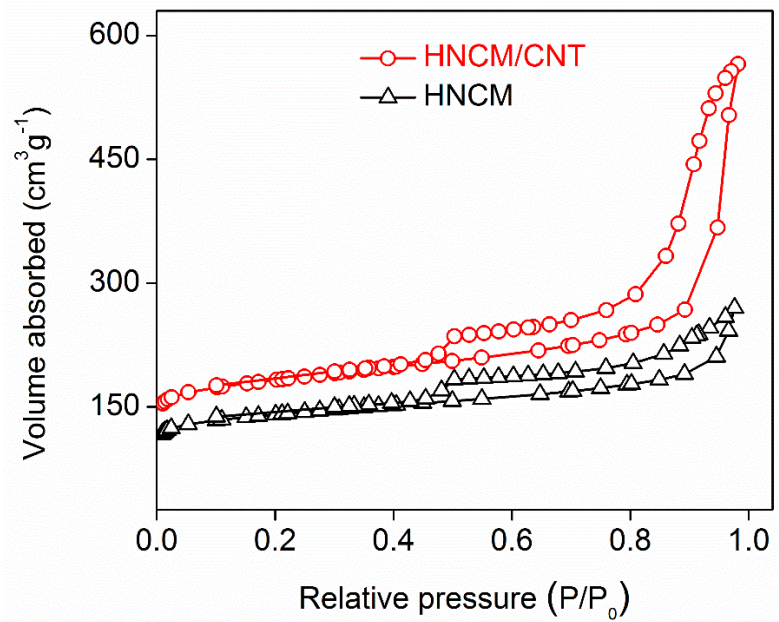

**Figure S7**. BET surface area of HNCM and HNCM/CNT.

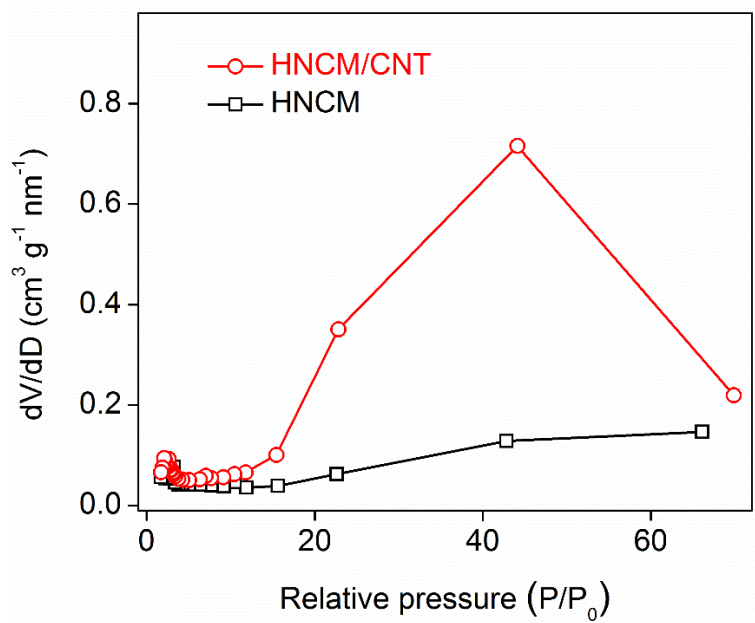

**Figure S8**. Pore size distribution of HNCM and HNCM/CNT.

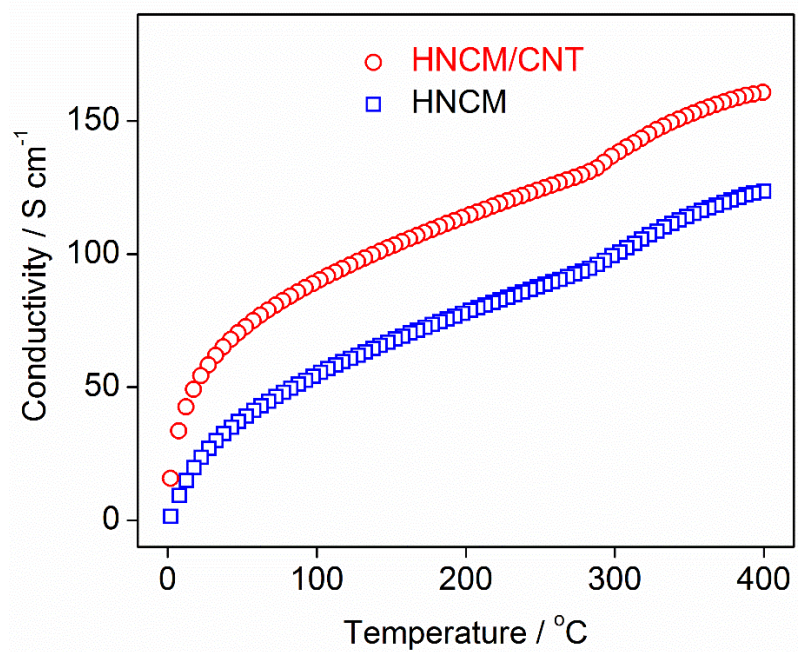

**Figure S9.** Temperature dependence of the conductivity measured for HNCM and HNCM/CNT from 5 K to 390 K using a four-probe method.

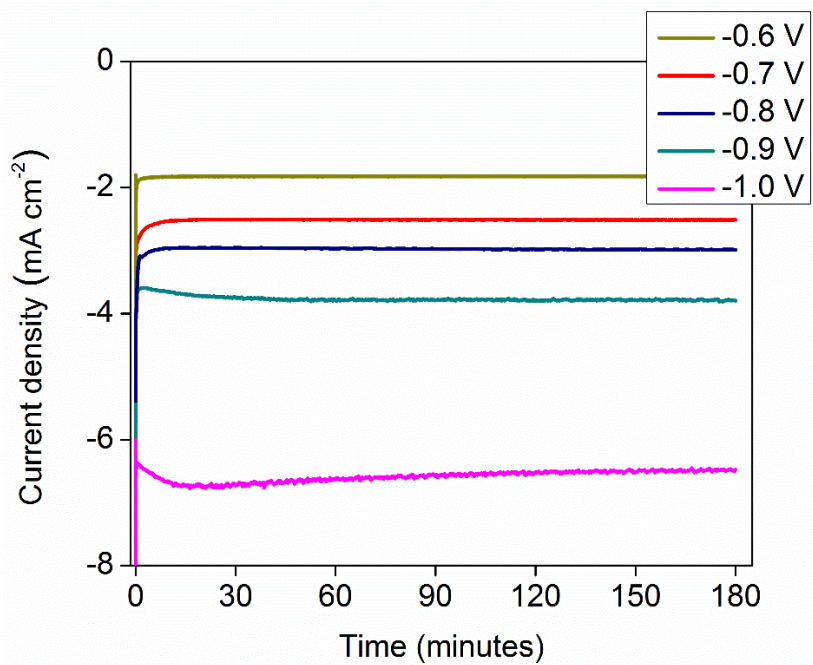

**Figure S10.** Time curves of the electrolysis process of HNCM electrode at different potentials (*vs* RHE).

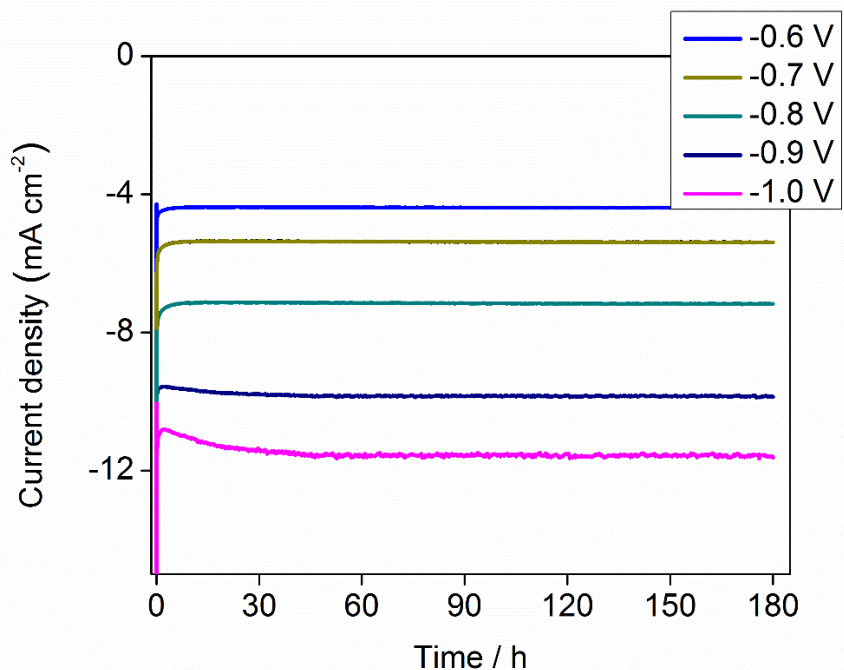

**Figure S11.** Time curves of the electrolysis process of HNCM/CNT electrode at different potentials (*vs* RHE).

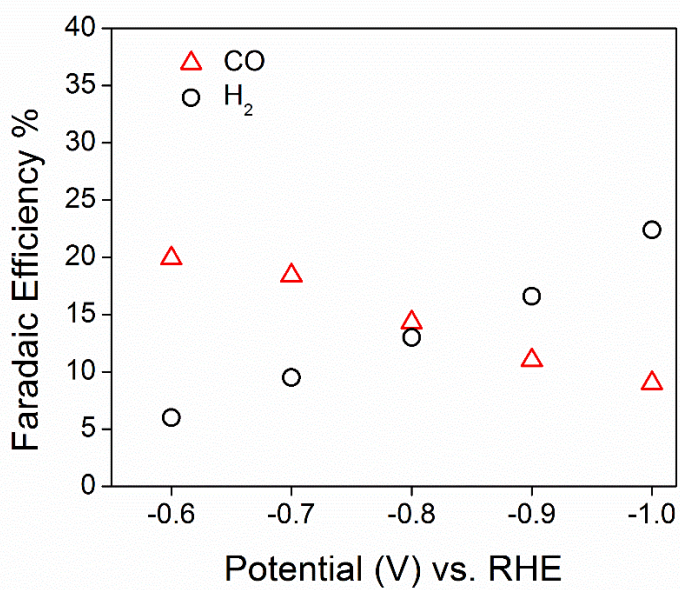

**Figure S12.** Faradaic efficiency of CO and $H_2$ generated on HNCM electrode.

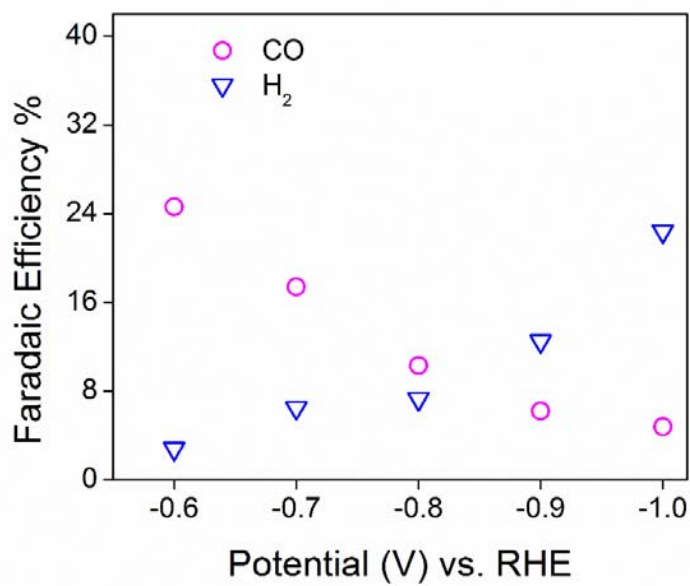

**Figure S13.** Faradaic efficiency of CO and $H_2$ generated on HNCM/CNT electrode.

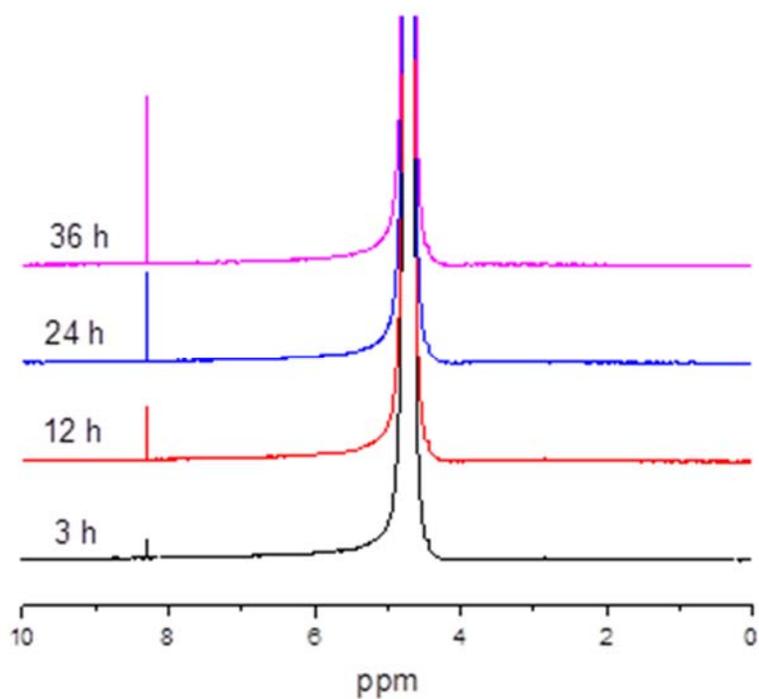

**Figure S14.** $^1$H-NMR spectra of electrolyte after $CO_2$ reduction at HNCM/CNT electrode under -0.9 V (*vs* RHE). The liquid samples for $^1$H-NMR testing were taken at different times during long-term operational stability testing.

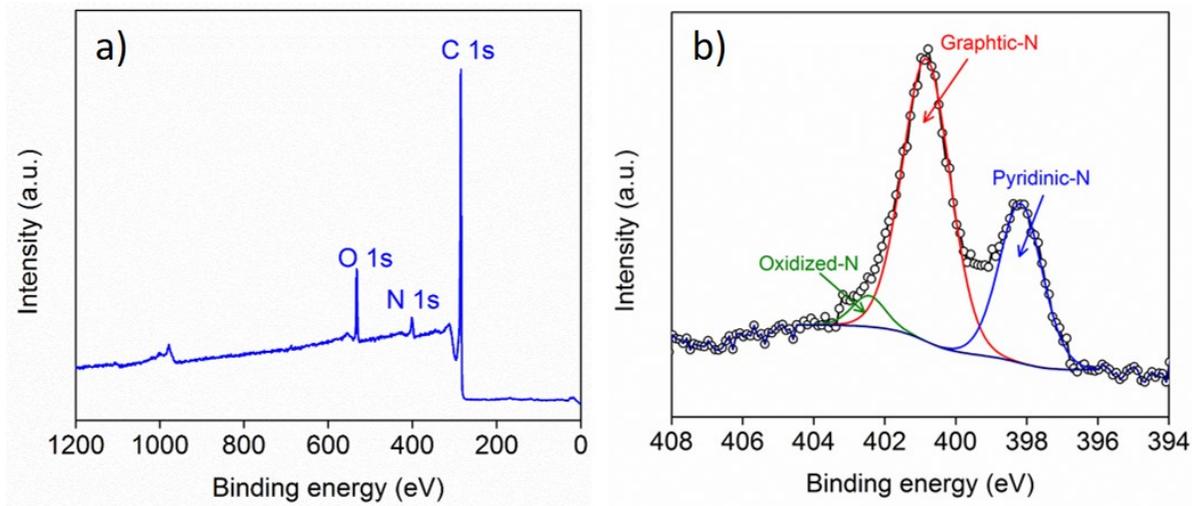

**Figure S15.** a) XPS spectrum of HNDCM/CNT after 36 h of electrolysis; b) High resolution N1 XPS spectra of HNDCM/CNT after 36 h of electrolysis. Similar to the original HNDCM/CNT, the N species in HNCM/CNT after 36 h of electrolysis exist in the form of graphitic N (56%) and pyridinic N (38%), except a slight fraction of oxidized N (6%) produced, indicating its relatively robust electrochemical stability.

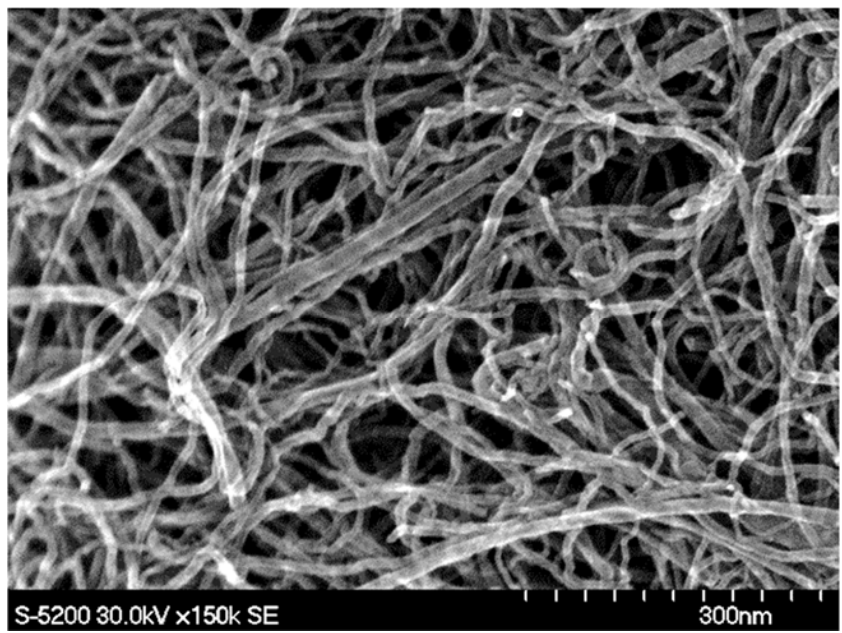

**Figure S16**. SEM image of CNTs after purified in 0.3M $HNO_3$ solution.

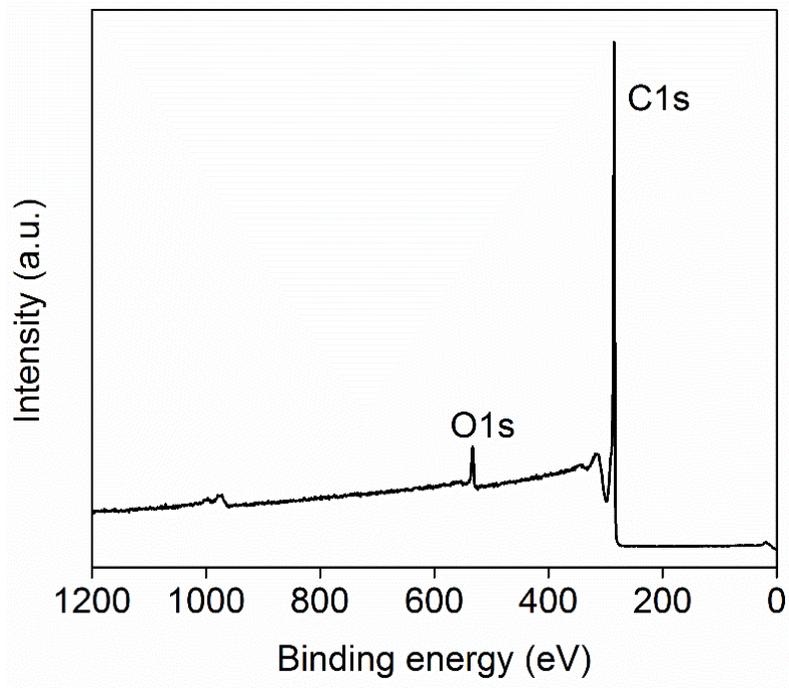

**Figure S17**. XPS spectra of purified CNTs. There are only C1s and O1s peaks in the XPS spectra, indicating metal-free character of purified CNTs.

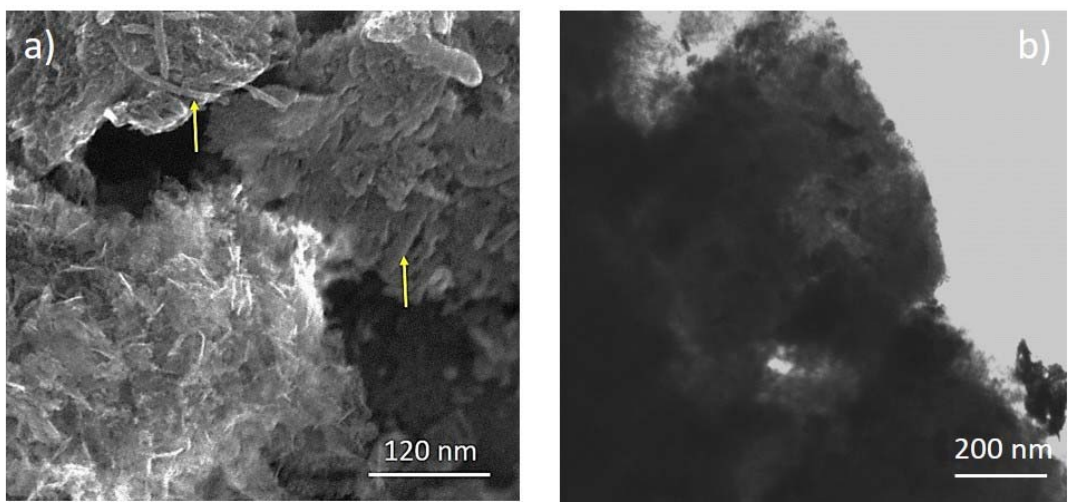

**Figure S18**. SEM image of N-doped carbon derived from CNT/PCMVImTf$_2$N. Yellow arrows indicate CNTs.

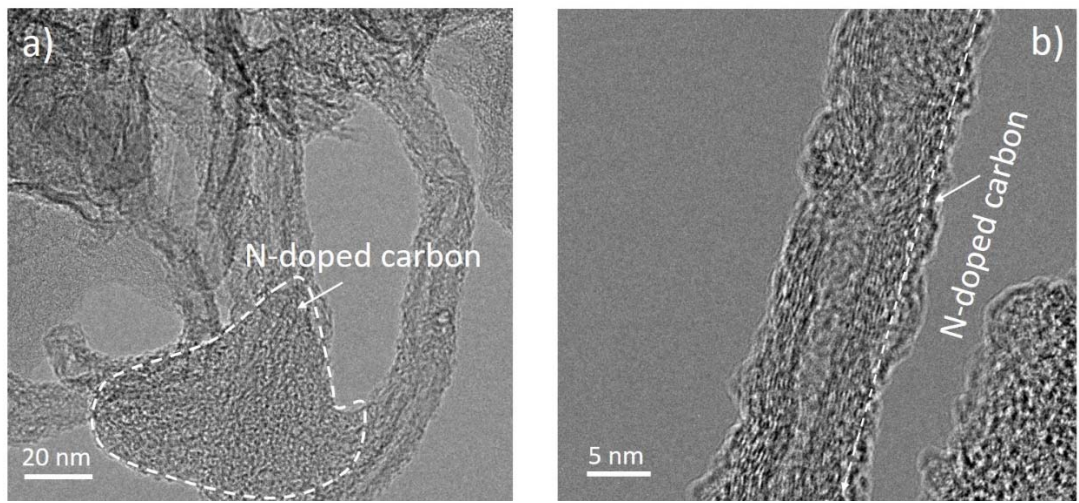

**Figure S19.** (a) TEM and (b) HRTEM images of N-doped carbon derived from CNT/PCMVImTf$_2$N. From Figure S19b, the core-shell structure of CNT-CN also can be observed, the formation mechanism of which is similar to that for HNDCM/CNT.

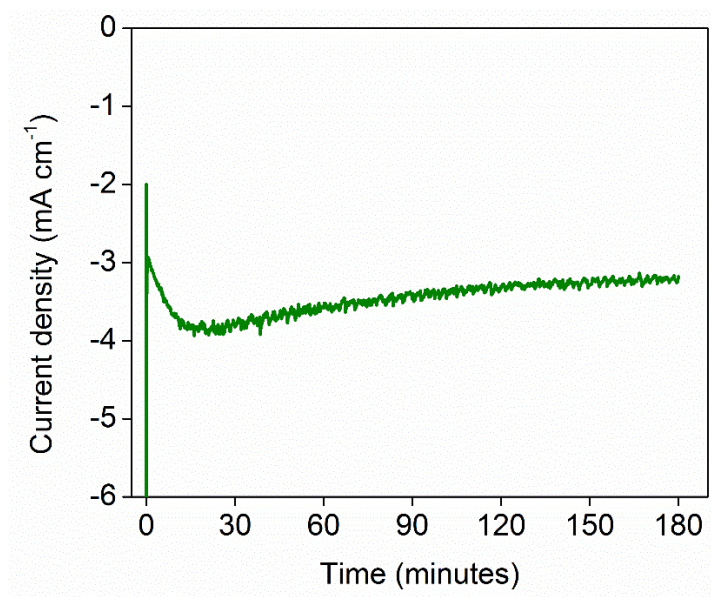

**Figure S20.** Time curve of the electrolysis process of N-doped carbon derived from CNT/PCMVImTf$_2$N electrode at -0.9V (*vs* RHE).